%% file: main.tex
\documentclass[journal]{IEEEtran}
\usepackage[nobreak]{cite}
\usepackage{graphicx}
\usepackage{float}
\usepackage{amsmath}
\usepackage{amssymb}
\usepackage{amsfonts}
\usepackage[printonlyused]{acronym}
\usepackage[capitalise]{cleveref}
\crefformat{equation}{#2(#1)#3}
\Crefformat{equation}{#2(#1)#3}
\crefmultiformat{equation}{#2(#1)#3}{ and~#2(#1)#3}{, #2(#1)#3}{, and~#2(#1)#3}
\usepackage{multirow}
\providecommand{\expe}[1]{\ensuremath{\mathrm{e}^{#1}}}

\usepackage{stackengine}
\newcommand\xrowht[2][0]{\addstackgap[.5\dimexpr#2\relax]{\vphantom{#1}}}
\def\j{\mathrm{j}}

\usepackage{url}
\usepackage{tikz}
\usetikzlibrary{calc}

\newcommand{\positiontextbox}[4][]{%
	\begin{tikzpicture}[remember picture,overlay]
		\node[inner sep=3pt, fill=yellow,align=left,draw,line width=1pt,#1] at ($(current page.north west) + (#2,-#3)$) {\parbox{.80\paperwidth}{#4}};
	\end{tikzpicture}%
}

\begin{document}
	
	\title{Direct Antenna Frequency-Hopped M-FSK Modulation with Time-Modulated Arrays}	
	\author{Roberto Maneiro-Catoira,~\IEEEmembership{Member,~IEEE,}
		Julio Br\'egains,~\IEEEmembership{Senior~Member,~IEEE,}\\
		Jos\'e A. Garc\'ia-Naya,~\IEEEmembership{Senior~Member,~IEEE,}
		and~Luis Castedo,~\IEEEmembership{Senior~Member,~IEEE}
		\thanks{%
			Manuscript received 15 October 2023; accepted 2 November 2023. Date
			of publication 6 November 2023; date of current version 5 February 2024.
			This work was supported in part by Grant ED431C 2020/15 funded by
			Xunta de Galicia and ERDF Galicia 2014-2020; in part by Grant PID2022-
			137099NB-C42 (MADDIE) and Grant TED2021-130240B-I00 (IVRY) funded
			by MCIN/AEI/10.13039/501100011033; in part by the European Union
			NextGenerationEU/PRTR; and funding for open access charge: Universidade
			da Coruña/CISUG. (Corresponding author: José A. García-Naya.)
			
			The authors are with the Universidade da Coruña (University of
			A Coruña), CITIC Research Center, 15071 A Coruña, Spain (e-mail: roberto.maneiro@udc.es; julio.bregains@udc.es; jagarcia@udc.es;
			luis.castedo@udc.es).
			
			Digital Object Identifier 10.1109/LAWP.2023.3330435
		}
	}
	
	\markboth{IEEE ANTENNAS AND WIRELESS PROPAGATION LETTERS, VOL. 23, NO. 2, FEBRUARY 2024}%
	{}
	\maketitle
	
	\input{acronymDef}
	\begin{abstract}
		We present an innovative approach that simultaneously enables direct antenna frequency hopped M-ary frequency shift keying (DAFH-MFSK) modulation and beamsteering through the use of time-modulated arrays (TMAs). The distinctive feature of our approach lies in the modulation of the TMA excitations with binary periodic sequences which can be easily frequency-adjusted and time-delayed to simultaneously allow for DAFH-MFSK direct antenna modulation and beamsteering. Notably, our TMA proposal offers a distinct advantage over conventional architectures in terms of performance metrics, including reduced insertion losses and enhanced phase resolution for beam steering, while also simplifying hardware complexity.
	\end{abstract}
	
	
	
	%
	
	\positiontextbox{11cm}{27cm}{\footnotesize This work is licensed under a Creative Commons Attribution 4.0 License. For more information, see \url{https://creativecommons.org/licenses/by/4.0/}. \\ The final published version of this work is available open access, with the same license, at \url{https://doi.org/10.1109/LAWP.2023.3330435}}

	\section{Introduction}
	\IEEEPARstart{L}{ow}-power and low-data-rate \ac{IoT} wireless devices widely use \ac{FSK} modulation due to their simplicity and resilience to noise and attenuation \cite{Keysight}. In standard \ac{MFSK} \cite[Chapter 5]{goldsmith2005wireless}, a constant-amplitude sine carrier with frequency $f_c{+}f^m_{\text{FSK}}=f_c{+}m\Delta f_{\text{FSK}}$, $m {\in} \mathcal{M}{=}\{1,2,\dots,M\}$ is selected every symbol period $T_s$, depending on which symbol is to be transmitted, being $f_c$ the reference (or base) carrier frequency, and $\Delta f_{\text{FSK}}$ the separation between two adjacent $f^m_{\text{FSK}}$ values. Accordingly, $B_{\text{FSK}}{=}M\Delta f_{\text{FSK}}$ corresponds to the total \ac{FSK} bandwidth.
	
\begin{figure}[!t]
		\centering
		\includegraphics[width=0.9\columnwidth]{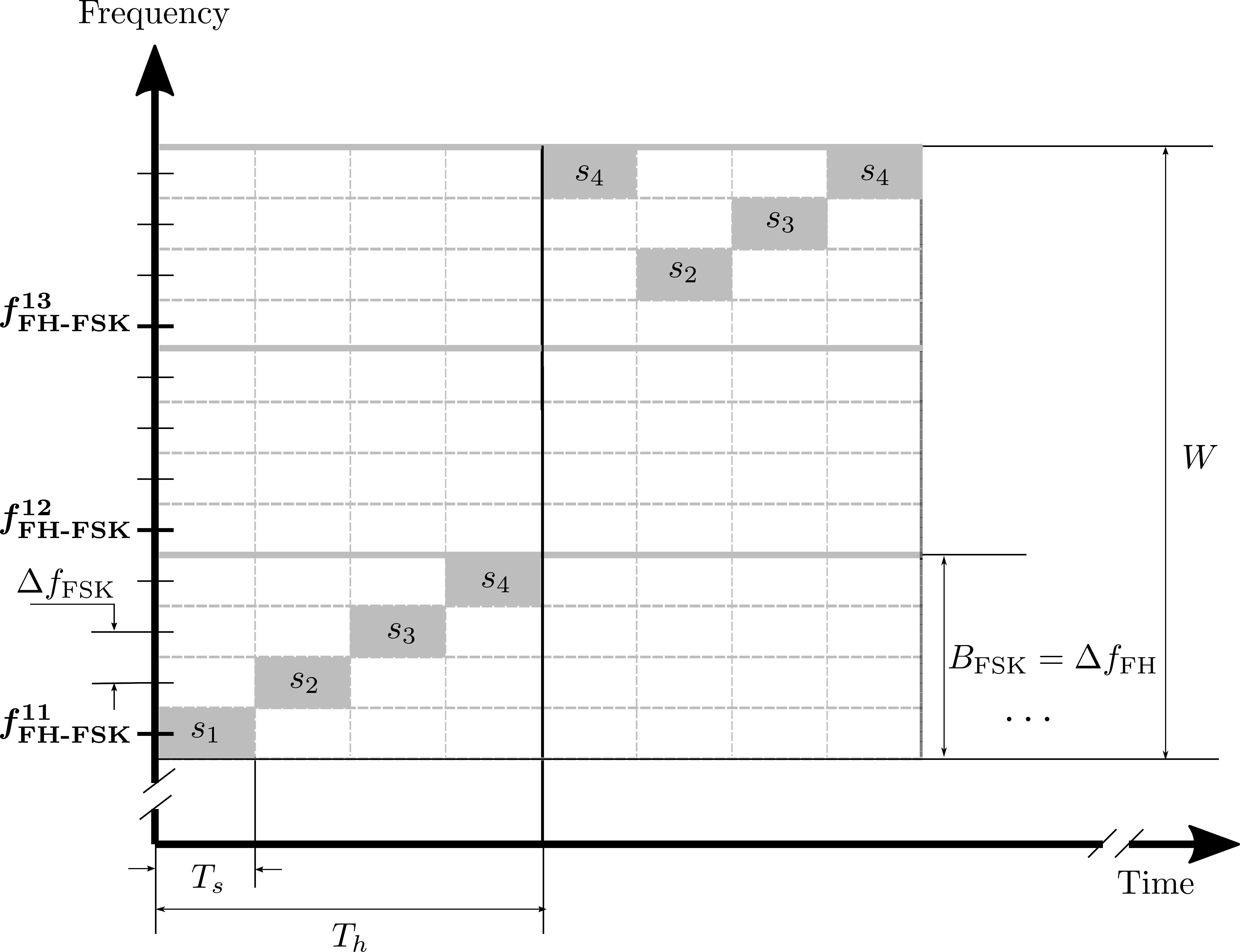}
		\caption{Time-frequency diagram for slow  \ac{FH}-\ac{FSK} modulation with $M{=}4$, $K{=}3$, and $L{=}4$. The transmitted \ac{MFSK} symbols are denoted by $s_m$, $m {\in} \mathcal{M}$.} 
		\label{fig:Time-frequency diagram}
	\end{figure}

	Despite their advantages, \ac{MFSK} signals can be easily intercepted and, moreover, can be seriously  distorted by frequency selective channels. These drawbacks can be overcome by means of \ac{FH} techniques \cite[Chapter 13]{goldsmith2005wireless}, which randomly change the carrier frequency in every hop period $T_h$. We assume hopping frequencies that are randomly selected from a set of equally spaced frequencies $f_{\text{FH}}^k{=}(k{-}1)\Delta f_{\text{FH}}$, $k {\in} \mathcal{K}{=}\{1,2,\dots,K\}$ with $\Delta f_{\text{FH}}$ being the separation between any two adjacent $f_{\text{FH}}^k$. We consider $\Delta f_{\text{FH}} {=} M\Delta f_{\text{FSK}}$, hence the transmit frequency for every $T_s$ in \ac{FH}-\ac{FSK} is $f_{\text{FH-FSK}}^{mk} {=} f_c {+} f_{\text{FSK}}^m {+} f^k_{\text{FH}} {=} f_c {+} [m {+} (k {-} 1)M]\Delta f_{\text{FSK}}$, whereas $W {=} K\Delta f_{\text{FH}} {=} KB_{\text{FSK}}$ is the total transmission bandwidth. We focus on slow \ac{FH} where $T_h {=} L T_s$ ($L {\in} \mathbb{N}$), as shown in \cref{fig:Time-frequency diagram}, which contains the time-frequency plot of an \ac{FH}-\ac{FSK} transmission. Since demodulation requires knowledge of the pseudo-random \ac{FH} pattern, \ac{FH}-\ac{FSK} prevents eavesdropping while increases robustness to frequency-selective channels. 
	\begin{figure}[!t]
		\centering
		\includegraphics[width=0.9\columnwidth]{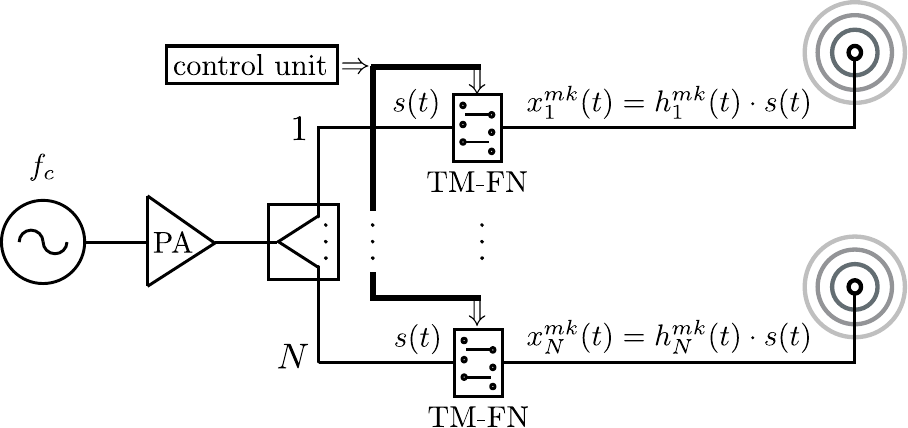}
		\caption{Proposed \ac{TMA} architecture to jointly perform \ac{DAFH-MFSK} modulation and beamsteering. The schematic of a \ac{TMFN} module is detailed in \cref{fig:TMFN}b). For each modulating waveform $h_n^{mk}(t)$, $n {\in} \mathcal{N}$ denotes the corresponding antenna element, $m {\in} \mathcal{M}$ is the \ac{FSK} symbol transmitted, and $k {\in} \mathcal{K}$ is the hop frequency slot. In the text, the values of $m$ and $k$ will be single digits so that the notation with the superscript $mk$ is not misleading.} 
		\label{fig:architecture}
	\end{figure}
	\begin{figure*}[ht]
		\centering
		\includegraphics[width=1.9\columnwidth]{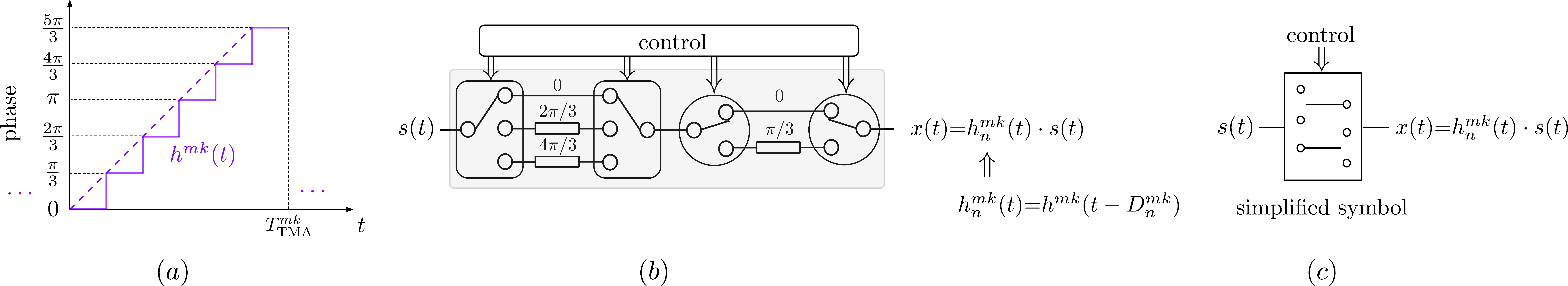}
		\caption{(a) Periodical ($T_{\text{TMA}}^{mk}$) six-level \ac{LP-TM} signal which allows for the direct transmission of the $m$-th \ac{FSK} symbol over the $k$-th \ac{FH} slot using a \ac{TMA}. Notice that only $\measuredangle h^{mk}(t)$ is sketched since $|h^{mk}(t)|{=}1$ $\forall m, k$. To additionally perform beamsteering, in the $n$-th antenna element, $h^{mk}(t)$ is subject to a time-delay $D^{mk}_n$, i.e., $h_n^{mk}(t) {=} h^{mk}(t {-} D_n^{mk})$. (b) Switched \ac{TMFN} of the $n$-th \ac{TMA} element, which consists of two \ac{SP3T} switches and two \ac{SPDT} switches or, equivalently, six \ac{SPDT} switches, to time modulate an input signal $s(t)$ with $h^{mk}_n(t)$; and (c) simplified block diagram of (b).} 
		\label{fig:TMFN}
	\end{figure*}
	
	Another concept, in line with \ac{IoT}, is \ac{DAM}, which consists in modulating the carrier in the antenna itself \cite{Fusco1999,Keller2006}. \ac{DAM} replaces baseband modulation and significantly reduces transmission \ac{HW} complexity, cost, and power consumption \cite{Henthorn2017,Henthorn2020,James2016}. Furthermore, since modulation occurs after amplification, \acp{PA} only need to amplify a single carrier, hence avoiding wideband \acp{PA}. 
	
	This letter combines \ac{FH}-\ac{FSK} and \ac{DAM} into a transmission method termed \ac{DAFH-MFSK} and proposes an innovative approach for its implementation using a highly efficient and versatile \acf{SSB}\footnote{\ac{SSB} \acp{TMA} remove unwanted frequency-mirrored beam patterns produced by conventional \acp{TMA} to achieve high efficiency levels.} \ac{TMA} \cite{Rocca2016,Maneiro2019a,Maneiro2019c,QChen2020,Ni2021TAP,Aksoy2012} which, in addition, is capable of performing beamsteering.

	\section{Joint DAFH-MFSK Modulation and Beamsteering: A TMA Approach}\label{Sec:Design}
	\subsection{\ac{SSB} Time Modulating Feeding Network \label{subsec:SSB}}
	
	\cref{fig:architecture} plots the proposed \ac{SSB} \ac{TMA} architecture for \ac{DAFH-MFSK}, equipped with periodic \acfp{TMFN} to jointly perform \ac{DAFH-MFSK} modulation and beam steering. We consider a linear array of $N$ isotropic elements with unitary static excitations, $I_n{=}1$, $n{\in} \mathcal{N}$, and $\mathcal{N}{=}\{1,\dots,N\}$. The $n$-th element excitation is modulated during a symbol period $T_s$ by the periodic pulsed signal $h_n^{mk}(t)$, which is a time-shifted version of the stair-step approximation to $h^{mk}(t)$ (see \cref{fig:TMFN}a), an \ac{LP-TM} waveform \cite{Ni2021TAP} with unit amplitude and phase varying from $0$ to $5\pi/3$ (six steps). Therefore, $h_n^{mk}(t) {=} h^{mk}(t {-} D_n^{mk})$, being $D_n^{mk}$ a variable time-delay. Notice that $h_n^{mk}(t)$ has a fundamental period $T_{\text{TMA}}^{mk} {=} 1/f_{\text{TMA}}^{mk} {\ll} T_s$, where $m {\in} \mathcal{M}$ refers to the transmitted $m$-th \ac{FSK}-modulated level, and $k {\in} \mathcal{K}$ accounts for the hop frequency slot selected during $T_s$. 
	
\begin{figure}[b]
		\centering
		\includegraphics[width=0.90\columnwidth]{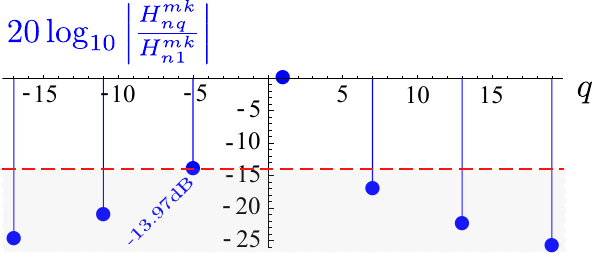}
		\caption{Normalized Fourier series power spectrum of $h_n^{mk}(t)$. All unwanted harmonics have a minimum rejection level of $13.97$\,dB with respect to the exploited harmonic at $q{=}1$. Notice that $\left|H^{mk}_{nq}\right|$ is the same for all values of $m$, $k$, and $n$ for a given $q$ because $h_n^{mk}(t)$ is either a time-scaled version (when $m$ and/or $k$ changes) or a time-shifted version (when $n$ changes) of the same waveform.} 
		\label{fig:Normalized Fourier}
	\end{figure}

	The synthesis of $h_n^{mk}(t)$ is described using a switched \ac{TMFN} (see \cref{fig:TMFN}b). Considering the rectangular pulse signal $p^{mk}(t){=}1$ when $0{\leq}t{<}T_{\text{TMA}}^{mk}/6$, and $p^{mk}(t){=}0$ otherwise, we can express a single period of $h^{mk}(t)$ as $\sum_{l=0}^5 p^{mk}(t {-} \tfrac{lT_{\text{TMA}}^{mk}}{6})e^{j\frac{2\pi l}{6}}$. Hence, the exponential Fourier series coefficients of the periodic signal $h^{mk}(t)$ are:
	\begin{align} 
		\label{eq:Fourier coefficients of h(t)}
		H_q^{mk}=\begin{cases} 
			\frac{6}{\pi q}\sin\left(\frac{\pi q}{6}\right)e^{j\frac{\pi q}{6}}, & q {\in} \Psi,\\
			0, & q {\notin} \Psi; \Psi {=} \{q {=} 6i {+} 1; i {\in} \mathbb{Z}\}
		\end{cases}
	\end{align}
	The Fourier coefficients $H_q^{mk}$ are the same for all values of $m$ and $k$ because $h^{mk}(t)$ is always a time-scaled version of the waveform shown in \cref{fig:TMFN}a. In view of \cref{eq:Fourier coefficients of h(t)}, the Fourier coefficients of 
	$h_n^{mk}(t) {=} h^{mk}(t {-} D_n^{mk})$ are given by
	\begin{align} 
		\label{eq:Fourier coefficients of h^{mk}(t)}
		H_{nq}^{mk}=\begin{cases} 
			H_q^{mk} e^{-j2\pi q f_{\text{TMA}}^{mk} D_n^{mk}}, &  q \in \Psi,\\
			0, &q \notin \Psi
		\end{cases}
	\end{align}
	and the exponential Fourier series expansion of $h_n^{mk}(t)$ is 
	\begin{equation}\label{eq:Fourier series}
		h_n^{mk}(t)=\sum_{q\in \Psi}H_q^{mk}\expe{-\j 2\pi q f_{\text{TMA}}^{mk} D_n^{mk}}\expe{\j 2\pi q f_{\text{TMA}}^{mk} t}.
	\end{equation}
	According to \cref{eq:Fourier coefficients of h^{mk}(t)}, \cref{fig:Normalized Fourier} shows the normalized Fourier series power spectrum of $h_n^{mk}(t)$ in dB, namely $20\log_{10}\left|{H^{mk}_{nq}}/{H^{mk}_{n1}}\right|$. We can see, apart from the \ac{SSB} property of the waveform, that the most meaningful unwanted harmonic is the one with order $q{=}-5$, whose relative level is at $-13.97$\,dB with respect to the useful harmonic $q{=}1$. 
	
	\ac{TMA} periodic modulating signals, $h_n^{mk}(t)$, have the following features: (1) they have no frequency-mirrored harmonics (hence the term \ac{SSB}) and their first positive harmonic concentrates almost all the transmitted energy; (2) this harmonic is located at $f_{\text{TMA}}^{mk} {=} f_{\text{FH}}^k {+} f^m_{\text{FSK}}$, hence the \ac{TMA} transmits the $m$-th level within the $k$-th \ac{FH} slot; and (3) the phase term of the first positive Fourier coefficient of  $h_n^{mk}(t)$  is proportional to the time delay $D_n^{mk}$ (see \cref{eq:Fourier coefficients of h^{mk}(t)}), which is instrumental to determine the steering direction of the \ac{TMA} beampattern and, unlike digital \acp{VPS} in standard phased arrays, $D_n^{mk}$ can be adjusted almost continuously \cite{QChen2019,Zeng2021}.
	
	In addition to the rejection threshold of unwanted harmonics shown in \cref{fig:Normalized Fourier}, the time-modulation efficiency of the \ac{TMA} is $\eta_{\text{TM}} {=}\mathcal{P}_U^{\text{TM}}/\mathcal{P}_R^{\text{TM}}$, where $\mathcal{P}_U^{\text{TM}}$ and $\mathcal{P}_R^{\text{TM}}$ are the respective useful and total average power radiated by the \ac{TMA}. According to \cite[Eq. 16]{Maneiro2017_a}, $\eta_{\text{TM}}$ is given by
	\begin{equation}\label{eq:efficiency_0}
		\eta_{\text{TM}} = {\left|H^{mk}_{n1}\right|^2}/{\sum_{q\in \Psi}\left|H^{mk}_{nq}\right|^2}.
	\end{equation}
	Since $h_n^{mk}(t)$ has unit amplitude, then
	\begin{equation}\label{eq:total energy}
		\frac{1}{T_{\text{TMA}}^{mk}}\int_0^{T_{\text{TMA}}^{mk}}|h_n^{mk}(t)|^2dt = \sum_{q \in \Psi}|H_{nq}^{mk}|^2 = 1,    
	\end{equation}
	thus obtaining (see \cref{eq:Fourier coefficients of h^{mk}(t)} and \cref{eq:efficiency_0})
	\begin{equation}\label{eq:efficiency}
		\eta_{\text{TM}} = |H_{n1}^{mk}|^2 = [6/\pi\cdot\sin(\pi/6)]^2 = 0.912.
	\end{equation}
	This means that the proposed \ac{SSB} \ac{TMA} architecture ensures that more than $91$\,\% of the total energy is transmitted over the first positive harmonic. On the basis of this result, the following approximation is applicable in \cref{eq:Fourier series}
	\begin{equation}\label{eq:h_n(t) tercera}
		h_n^{mk}(t)\approx H_1^{mk}\expe{-\j 2\pi f_{\text{TMA}}^{mk} D_n^{mk}}\expe{\j 2\pi f_{\text{TMA}}^{mk} t}
	\end{equation}

	\subsection{Signal Radiated During a Symbol Period}\label{subsec:Tx}
	
	As shown in \cref{fig:architecture}, the input to the proposed \ac{SSB} \ac{TMA} is the single-frequency carrier signal $s(t) {=} \expe{\j 2\pi f_c t}$. During a symbol period $T_s$, the \ac{TMA} excitations are time modulated  by $h_n^{mk}(t)$ and the signal radiated by the \ac{TMA} in the spatial direction $\theta$ is given by
	\begin{equation}\label{eq:Signal radiated}
		x^{mk}(\theta,t)= \sum_{n=1}^{N}h_n^{mk}(t)\expe{\j \beta_c z_n\sin\theta}s(t)
	\end{equation}
	where $z_n$ is the position of the $n$-th array element on the $z$ axis and $\beta_c {=} 2\pi/\lambda_c$ is the wavenumber for a carrier wavelength $\lambda_c {=} \mathrm{c} /f_c$. Normalizing \Cref{eq:h_n(t) tercera} with respect to $H_1^{mk}$, which is the same for all values of $m$ and $k$ (see \cref{eq:Fourier coefficients of h(t)}), Eq. \Cref{eq:Signal radiated} can be rewritten as
	\begin{align}\label{eq:Signal radiated aprox}
		x^{mk}(\theta,t) &\approx \sum_{n=1}^{N} \expe{\j 2\pi\left(\frac{z_n}{\lambda_c} \sin\theta {-} f_{\text{TMA}}^{mk} D_n^{mk}\right)}\expe{\j 2\pi \left(f_c {+} f_{\text{TMA}}^{mk}\right) t}\notag\\
		&\approx \underbrace{\mathrm{AF}^{mk}(\theta)}_{\text{BS}}\hspace{0.5cm}\cdot \underbrace{\vphantom{\mathrm{AF}^{mk}(\theta)} \expe{\j 2\pi f_{\text{FH-FSK}}^{mk}t}}_{\text{DAFH-MFSK modulation}}
	\end{align}
	where the term $\mathrm{AF}^{mk}(\theta) {=} \sum_{n=1}^{N} \expe{\j 2\pi\left(\frac{z_n}{\lambda_c} \sin\theta {-} f_{\text{TMA}}^{mk} D_n^{mk}\right)}$ is the spatial array factor during $T_s$ and provides the beamsteering ability to the \ac{TMA}. Indeed, the maximum of the radiation pattern can be pointed to the direction $\theta_0$ by adjusting the delays $D_n^{mk}$, $n {\in} \mathcal{N}$, so that the following equation is satisfied
	\begin{equation}\label{eq:calculo delays}
		\expe{\j 2\pi f_{\text{TMA}}^{mk} D_n^{mk}} = \expe{\j 2\pi\frac{z_n}{\lambda_c} \sin\theta_0}
	\end{equation}
	in which case the array factor is
	\begin{equation}
		\mathrm{AF}^{mk}(\theta)=\sum_{n=1}^{N} \expe{\j 2\pi\frac{z_n}{\lambda_c}\left( \sin\theta - \sin\theta_0\right)} 
	\end{equation}
	On the other hand, the term $\expe{\j 2\pi f_{\text{FH-FSK}}^{mk}t}$ in \cref{eq:Signal radiated aprox} allows the \ac{TMA} to transmit the $m$-th \ac{FSK} level over the $k$-th \ac{FH} slot, and thus perform \ac{DAFH-MFSK} modulation.
	
	\begin{figure}[ht]
		\centering
		\includegraphics[width=0.90\columnwidth]{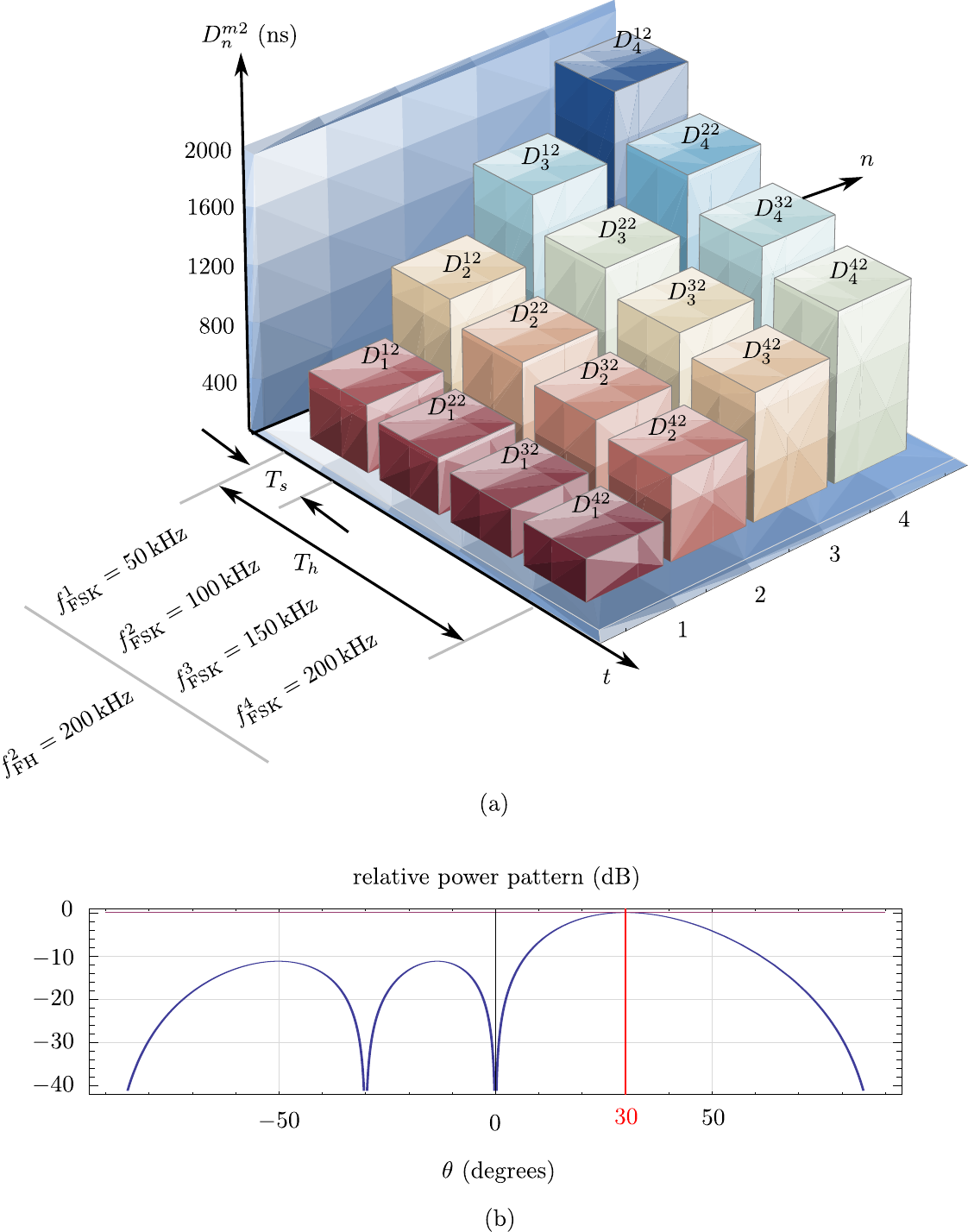}
		\caption{(a) Time delays $D_n^{m2}$ specified in \cref{tab:tableI} as a function of $n {\in} \mathcal{N}$ and time. During $T_h$, the sequence of symbols $s_m$, corresponding to the frequencies $f_{\text{FSK}}^m$, $m {=} \{1,2,3,4\}$, are transmitted. (b) Relative power radiated pattern of the proposed \ac{TMA}. When its modulating waveforms are subject to $D_n^{m2}$ during the considered $T_h$, the beampattern points towards $\theta_0{=}30^{\circ}$.} 
		\label{fig:f_HP^2 30 degrees}
	\end{figure}
	
	\section{Case Study and Comparative Analysis}
	This section has a twofold purpose: (1) Demonstrate the feasibility of the proposed technique by means of numerical simulations, and (2) compare it with conventional architectures performing FH-MFSK and beamsteering simultaneously.
	\subsection{Numerical Example}\label{subsec:Simulations}
	
	Let us consider a \ac{DAFH-MFSK} modulator architecture based on \acp{TMFN} (refer to \cref{fig:architecture}). We assume the following parameters: carrier frequency: $f_c {=}2.5$\,GHz; number of antenna elements: $N{=}4$, spaced $\lambda_c/2$ apart; modulation scheme: 4-FSK ($M{=}4$); hopping frequencies: 6 possible values ($K{=}6$); symbol period: $T_s{=}10$\,ms; hop duration: $L{=}4$, thus $T_h {=} 4T_s {=} 4$\,ms; frequency separation for \ac{FSK}: $\Delta f_{\text{FSK}} {=} 50$\,kHz, with $f^1_{\text{FSK}} {=} 50$\,kHz; and frequency spacing for \ac{FH}: $\Delta f_{\text{FH}} {=} 200$\,kHz, with $f^1_{\text{FH}} {=} 0$\,Hz. The minimum frequency of the \ac{TMA} modulating waveforms is given by $f_{\text{TMA}}^{11} {=} f_{\text{FH}}^1 {+} f^1_{\text{FSK}} {=} 50$\,kHz, corresponding to their maximum possible period $T_{\text{TMA}}^{\text{max}} {=} T_{\text{TMA}}^{11} {=} 1/f_{\text{TMA}}^{11} {=} 20\,\text{\textmu s} \ll T_s {=} 10$\,ms. Additionally, the \ac{TMA} offers a minimum period of $T_{\text{TMA}}^{\text{min}} {=}  T_{\text{TMA}}^{46} {=} 1/f_{\text{TMA}}^{46} {=} 833$\,ns.
	
	As an example, let us consider that the \ac{TMA} radiates sequentially the four 4-\ac{FSK} levels $m {=} \{1,2,3,4\}$ over the second ($k{=}2$) \ac{FH} slot of duration $T_h$ and centered at frequency $f_{\text{FH}}^2 {=} f_{\text{FH}}^1 {+} \Delta f_{\text{FH}} {=} 250$\,kHz. The frequencies to be transmitted and the antenna elements time delays $D_n^{m2}$, $n {\in} \mathcal{N}$,  (determined according to \cref{eq:calculo delays}) to point the maximum of the radiation pattern towards $\theta_0 {=} 30^{\circ}$ are shown in \cref{fig:f_HP^2 30 degrees} and summarized in \cref{tab:tableI}.
	
	
	\begin{figure}[ht]
		\centering
		\includegraphics[width=1.0\columnwidth]{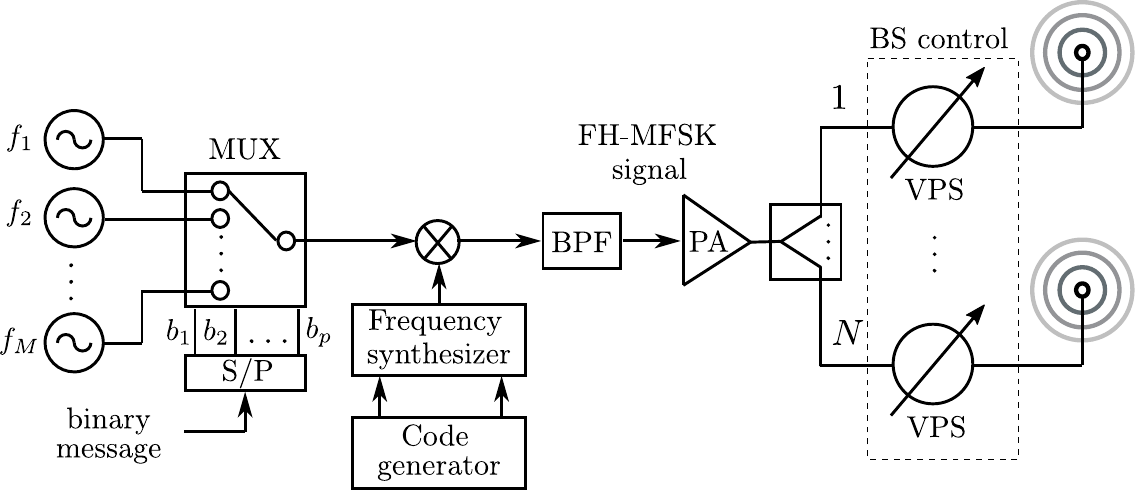}
		\caption{Block diagram of a conventional FH-MFSK transmitter followed by a standard phased array with digitally tuned passive \ac{VPS} to perform beamsteering.}   
		\label{fig:conventional architecture}
	\end{figure}
	
	\begin{table}[b]
		\caption{Frequencies and time delays $D_n^{m2}$ (nanoseconds) of the \ac{TMA} modulating waveforms for the hop frequency $f_{\text{FH}}^2{=}200\,\mathrm{kHz}$ and the four 4-\ac{FSK} levels $m=\{1,2,3,4\}$ to steer the beam towards $\theta_0 {=} 30^{\circ}$ (see \cref{fig:f_HP^2 30 degrees}) }
		\label{tab:tableI}
		\centering
		\begin{tabular}{@{}r@{\hspace{3pt}}|@{\hspace{3pt}}c@{\hspace{3pt}}|@{\hspace{3pt}}c@{\hspace{3pt}}|@{\hspace{3pt}}c@{\hspace{3pt}}|@{\hspace{3pt}}c@{\hspace{3pt}}|@{\hspace{3pt}}c@{\hspace{3pt}}}
			$\boldsymbol{m}$ & $\boldsymbol{f_{\text{FSK}}^m {+} f_{\text{FH}}^2}$ & $\boldsymbol{D_1^{m2}}$ & $\boldsymbol{D_2^{m2}}$ & $\boldsymbol{D_3^{m2}}$ & $\boldsymbol{D_4^{m2}}$\xrowht[()]{8pt}\\
			\hline\hline\xrowht[()]{6pt}
			1&\text{250\,kHz} & 500 & 1000 & 1500 & 2000\\\xrowht[()]{6pt}
			2&\text{300\,kHz}  & 417 & 833 & 1250 & 1667\\\xrowht[()]{6pt}
			3&\text{350\,kHz} & 357 & 714 & 1071 & 1429\\\xrowht[()]{6pt}
			4&\text{400\,kHz} & 313 & 625 & 938 & 1250\\
		\end{tabular}
	\end{table}
	
	\subsection{Comparison with Conventional Techniques}\label{subsec:Comparison}
	
	\cref{fig:conventional architecture} shows the block diagram of a conventional FH-MFSK transmitter followed by a standard phased array equipped with digitally tuned passive \acp{VPS} to perform beamsteering. Every $T_s$, the incoming binary data bits are employed, via a \ac{MUX}, to select the transmitting carrier frequency from a pool of $M$ possibilities. The FH-MFSK signal is generated by mixing the resulting \ac{MFSK} modulated signal with a carrier obtained from a digital frequency synthesizer under the control of a code generator. 
	
	Given that the mixer produces both sum and difference frequency components, but only the sum frequency is intended for radiation, a \ac{BPF} is placed after the mixer. Following amplification through the \ac{PA}, the signal is radiated to a given direction by adjusting the \acp{VPS} within the standard phased array. 
	
\begin{figure}[ht]
		\centering
		\includegraphics[width=0.95\columnwidth]{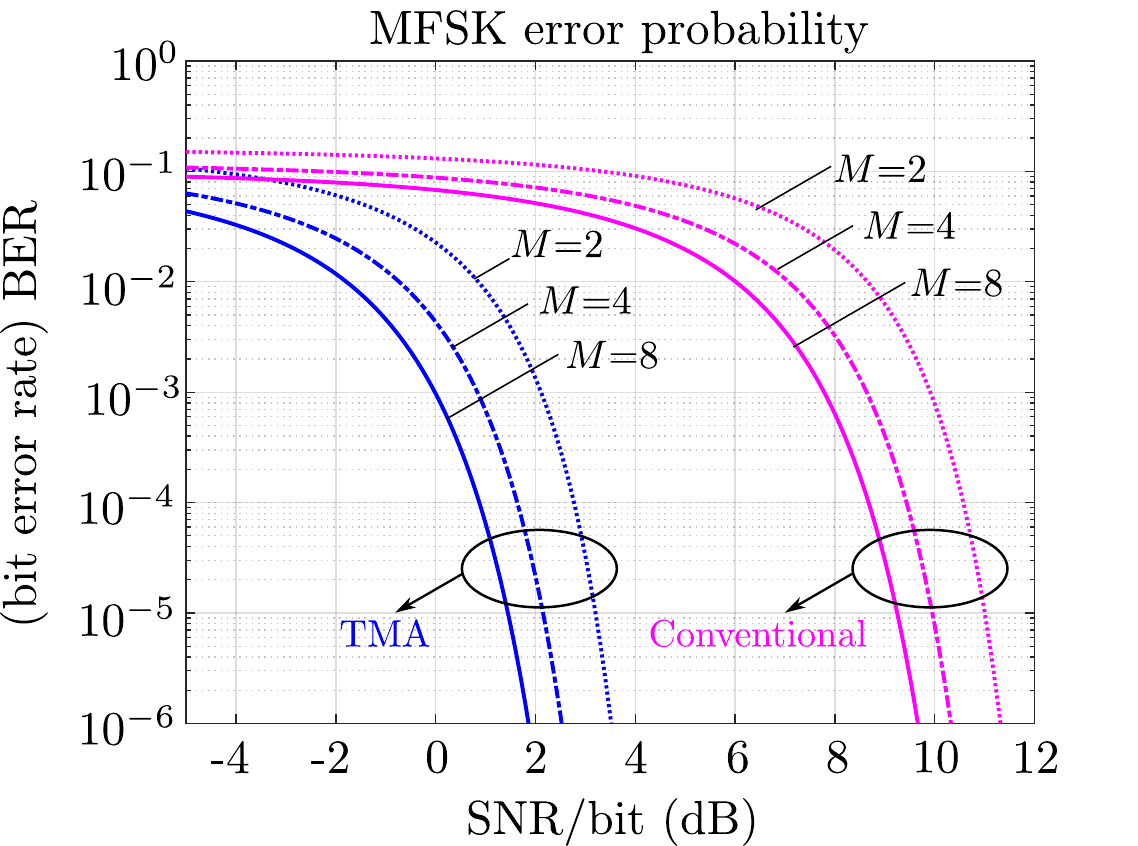}
		\caption{Performance of \ac{MFSK} demodulation for the proposed \ac{TMA} approach (see \cref{fig:architecture}) and the conventional one (see \cref{fig:conventional architecture}) in terms of \ac{BER} versus received \ac{SNR} per bit in the conventional scheme.}
		\label{fig:BER Comparison}
	\end{figure}

	Compared with the conventional scheme in \cref{fig:conventional architecture}, our proposed \ac{TMA} approach in \cref{fig:architecture} offers several key advantages:
	\begin{enumerate}
		\item Reduced hardware complexity: In our approach, there is no need for a \ac{MUX}, mixer, frequency synthesizer, or \ac{BPF}, and the \ac{PA} only needs to amplify a single carrier.
		
		\item Minimal oscillator requirements: Unlike the conventional scheme illustrated in \cref{fig:conventional architecture}, which necessitates multiple oscillators corresponding to the number of \ac{MFSK} levels, $M$, our approach requires just one oscillator. Refer to \cref{tab:tableII} where we employ big $O$ notation \cite{Cormen2001} for a comprehensive hardware comparison.
		
		\item Reduced number of \ac{SPDT} switches: The number of \ac{SPDT} swtches in our approach is only half that of the conventional scheme employing 6-bit \acp{VPS} (minimum resolution comparable to that of the \ac{TMA} \cite{QChen2019}). This results in a linear complexity, as detailed in \cref{tab:tableII}.
		
		\item Improved insertion losses: For both the conventional architecture and the TMA approach, we consider off-the-shelf devices with the lowest insertion loss, denoted as $\eta$, within the specified frequency band. Specifically, $\eta_{\text{MUX}}{=}0.7$\,dB (\ac{SP4T} switch) \cite{Analog}, $\eta_{\text{mixer}}{=}4.5$\,dB \cite{Marki}, $\eta_{\text{BPF}}{=}2$\,dB \cite{Minicircuits}, $\eta_{\text{VPS}}{=}4$\,dB \cite{Analog}, $\eta_{\text{SPDT}}{=}0.5$\,dB \cite{Analog}, and ideal $1{:}N$ power splitters are taken into account. 
		
		Under these circumstances, the conventional architecture exhibits an insertion loss of $\eta_{\text{conv}} = \eta_{\text{MUX}}+ \eta_{\text{mixer}} + \eta_{\text{BPF}} + \eta_{\text{VPS}} = 11.2$\,dB. In contrast, the \ac{TMA} approach \cite{Maneiro2020a} demonstrates significantly lower insertion losses, with $\eta_{\text{TMA}} = -10\log_{10}\eta_{\text{TM}}+ 6\eta_{\text{SPDT}} = 3.4$\,dB. Consequently, assuming equal power levels in the respective carrier signals, the same \ac{PA} gain, and that the performance of all components remains consistent across the entire bandwidth, the proposed \ac{TMA} approach achieves a substantial insertion loss reduction of $\Delta \eta {=} \eta_{\text{TMA}} {-} \eta_{\text{conv}} {=} -7.8$ dB. This insertion loss reduction leads to a significant performance improvement of \ac{MFSK} demodulation in terms of \ac{BER} versus received \ac{SNR} per bit, as illustrated in \cref{fig:BER Comparison}.
	\end{enumerate}
	
	\begin{table}[t]
		\caption{Comparative advantages of the proposed \ac{TMA} approach.} 
		\label{tab:tableII}
		\centering
		\setlength\tabcolsep{0.2em}
		\def\arraystretch{1.5}
		\begin{tabular}{| c | c | c | c |}
			\hline
			{FH-MFSK \& BS Scheme}&{\# Oscillators}&{\# SPDT switches} & {$\eta$\,(dB)}\\
			\hline \hline
			Conventional (\cref{fig:conventional architecture}) & $M\in O(M)$ &  $12N\in O(N)$ & 11.2\\ \hline  
			This work: \ac{TMA} approach (\cref{fig:architecture}) & $1\in O(1)$ &  $6N\in O(N)$ & 3.4\\\hline    
		\end{tabular}
	\end{table}
	
	\section{Conclusions}
	We have introduced an innovative \ac{TMA} approach that seamlessly combines \ac{DAFH-MFSK} modulation and \ac{BS}, making it particularly well suited for low-power and low-data-rate applications. This approach offers several key advantages over  comparable existing architectures, including better energy efficiency, simplified design, and the ability to achieve continuous phase-sensitivity beamsteering.
	\ifCLASSOPTIONcaptionsoff
	\newpage
	\fi
	
	\vfill
	
	
	
	\bibliographystyle{IEEEtran}
	%
	\bibliography{IEEEabrv,references}
	
	%

	
	\vfill
	

\end{document}

%% file: acronymDef.tex
\acrodef{ADC}[ADC]{analog-to-digital converter}
\acrodef{AF}[AF]{array factor}
\acrodef{AWGN}[AWGN]{additive white Gaussian noise}
\acrodef{ASK}[ASK]{amplitude-shift keying}
\acrodef{BER}[BER]{bit error ratio}
\acrodef{BF}[BF]{beamforming}
\acrodef{BS}[BS]{beamsteering}
\acrodef{BFN}[BFN]{beamforming network}
\acrodef{BPF}[BPF]{bandpass filter}
\acrodef{DAC}[DAC]{digital-to-analog converter}
\acrodef{DAM}[DAM]{direct antenna modulation}
\acrodef{DAFH-MFSK}[DAFH-MFSK]{direct antenna frequency hopped M-ary frequency shift keying}
\acrodef{DC}[DC]{direct current}
\acrodef{DFT}[DFT]{discrete Fourier Transform}
\acrodef{DOA}[DOA]{direction of arrival}
\acrodef{DSB}[DSB]{double sideband}
\acrodef{ETMA}[ETMA]{enhanced time-modulated array}
\acrodef{FDA}[FDA]{frequency diverse array}
\acrodef{FBW}[FBW]{fractional bandwidth}
\acrodef{FH}[FH]{frequency-hopping}
\acrodef{FSK}[FSK]{frequency-shift keying}
\acrodef{MFSK}[MFSK]{M-ary FSK}
\acrodef{FT}[FT]{Fourier Transform}
\acrodef{HDWT}[HDWT]{Haar Discrete Wavelet Transform}
\acrodef{HPBW}[HPBW]{half power beamwidth}
\acrodef{HT}[HT]{Hilbert Transform}
\acrodef{HW}[HW]{hardware}
\acrodef{IoT}[IoT]{internet of things}
\acrodef{IIoT}[IIoT]{industrial internet of things}
\acrodef{ISI}[ISI]{inter-symbol interference}
\acrodef{LMD}[LMD]{linearly modulated digital}
\acrodef{LNA}[LNA]{low noise amplifier}
\acrodef{LP-TM}[LP-TM]{linear-phase time-modulating}
\acrodef{MBPA}[MBPA]{Multibeam phased-array antenna}
\acrodef{NFC}[NFC]{Near Field Communication}
\acrodef{MMIC}[MMIC]{monolithic microwave integrated circuit}
\acrodef{MSLL}[MSLL]{maximum side-lobe level}
\acrodef{MSLFO}[MSLFO]{modified symmetric logarithmically increasing frequency offset}
\acrodef{MUX}[MUX]{multiplexer}
\acrodef{NMLW}[NMLW]{normalized main-lobe width}
\acrodef{NPD}[NPD]{normalized power density}
\acrodef{MRC}[MRC]{maximum ratio combining}
\acrodef{PA}[PA]{power amplifier}
\acrodef{PCB}[PCB]{printed circuit board}
\acrodef{PS}[PS]{phase shifter}
\acrodef{PSO}[PSO]{particle swarm optimization}
\acrodef{PSK}[PSK]{phase-shift keying}
\acrodef{PWM}[PWM]{pulse width modulation}
\acrodef{QAM}[QAM]{quadrature amplitude modulation}
\acrodef{RF}[RF]{radio frequency}
\acrodef{RFC}[RFC]{Rayleigh fading channel}
\acrodef{RFID}[RFID]{Radio Frequency Identification}
\acrodef{RPDC}[RPDC]{reconfigurable power/divider combiner}
\acrodef{SA}[SA]{simulated annealing}
\acrodef{SBR}[SBR]{sideband radiation}
\acrodef{SER}[SER]{symbol error rate}
\acrodef{SLL}[SLL]{sideband-lobe level}
\acrodef{SPMT}[SPMT]{single-pole multiple-throw}
\acrodef{SPST}[SPST]{single-pole single-throw}
\acrodef{SNR}[SNR]{signal-to-noise ratio}
\acrodef{SPDT}[SPDT]{single-pole dual-throw}
\acrodef{SP3T}[SP3T]{single-pole triple-throw}
\acrodef{SP4T}[SP4T]{single-pole four-throw}
\acrodef{SLFO}[SLFO]{symmetric logarithmically increasing frequency offset}
\acrodef{SSB}[SSB]{single sideband}
\acrodef{SSB-TM-MBPA}[SSB TM-MBPA]{single sideband time-modulated multibeam phased-array antenna}
\acrodef{SWC}[SWC]{sum of weighted cosines}
\acrodef{STA}[STA]{static array}
\acrodef{TM}[TM]{time-modulated}
\acrodef{TMFN}[TM-FN]{time-modulating feeding network}
\acrodef{TMA}[TMA]{time-modulated array}
\acrodef{TM-MBPA}[TM-MBPA]{time-modulated multibeam phased-array antenna}
\acrodef{VVA}[VVA]{variable-voltage attenuator}
\acrodef{ULA}[ULA]{uniformly-excited linear array}
\acrodef{UWB}[UWB]{ultra-wide band}
\acrodef{VPS}[VPS]{variable phase shifter}
\acrodef{VGA}[VGA]{variable-gain amplifier}